\documentclass{ws-ijmpa}

\begin{document}

\title{THE CASIMIR EFFECT AND THE FOUNDATIONS\\
OF STATISTICAL PHYSICS}

\author{V. M. MOSTEPANENKO$^{1,2}$ and
G. L. KLIMCHITSKAYA$^{1,3}$}

\address{${}^1$Institute for Theoretical
Physics, Leipzig University, Postfach 100920,
D-04009, Leipzig, Germany \\
${}^2${Noncommercial Partnership ``Scientific Instruments'',
Tverskaya St. 11, Moscow,
103905, Russia}\\
${}^3${North-West Technical University,
Millionnaya St. 5, St.Petersburg,
191065, Russia}
}
\maketitle
\begin{abstract}
The Lifshitz theory and its modifications are discussed with
respect to the Nernst heat theorem and the experimental data
of several recent experiments. An analysis of all available
information leads to the conclusion that some concepts of
statistical physics might need reconsideration.
\end{abstract}



\section{Introduction}

In the last few years there has been an explosion of interest in
the Casimir effect\cite{1} which has resulted in new precise
experiments, elaboration of powerful theoretical methods and in
suggestions of prospective applications (a modern overview of the
subject can be found in Ref.~\refcite{2}). Coincident with many
developments of a conclusive character, starting from 2000 there
were also controversial discussions in the literature on the
nature and size of the thermal effects in the Lifshitz theory of the
Casimir force.\cite{2,3} Bostr\"{o}m and Sernelius\cite{4}
were the first who predicted the existence of large thermal
corrections to the Casimir force between two plane parallel
metallic plates described by the Drude model spaced at separation
of a few hundred nanometers. Bordag et al.\cite{5} argued that
such corrections are nonphysical and suggested to calculate the
thermal Casimir force using the dielectric permittivity of the
plasma model (for the latter purpose the plasma model was also
used in Ref.~\refcite{6}). Later both approaches were further
developed in Refs.~\refcite{7,8} and
\refcite{9,10}, respectively. A step of paramount
importance was made by the experiments of Decca et
al.\cite{11,12} which excluded the existence of
large thermal corrections predicted by the Drude model at
almost 100\% confidence level. Related experiments for
semiconductor\cite{13} and dielectric\cite{14} materials leading
to similar conclusions\cite{15} were subsequently performed.
On the one hand, thermodynamic arguments based on the Nernst
heat theorem favored the plasma model approach for
metals\cite{10} and neglect of the dc conductivity for
dielectrics.\cite{16}
On the other hand, statistical physics applied in the so-called
classical limit was in support of the Drude model.\cite{17}
The situation was so extraordinary that it was even
suggested\cite{18}\cdash\cite{20} to modify the Lifshitz theory
providing the fundamental description of both the van der Waals
and Casimir forces between real materials. For this purpose
the standard reflection coefficients were replaced with their
generalizations taking into account the screening effects and
diffusion currents. It was shown, however, that the modified
theory still violates the Nernst heat theorem\cite{21}\cdash\cite{26}
and is in contradiction with the experimental data. These
conclusions were disputed\cite{27}\cdash\cite{29} by the authors of
the modified theory.

Keeping in mind that controversial discussion on this subject
has lasted
for already ten years and consensus is not yet achieved,
it seems pertinent to collect and analyze all the proposed
arguments.  Such an analysis seems to be especially useful
because there were discussions in the previous literature which
appear one sided by dealing with only selected facts and
disregarding others. By taking into account all known facts
in a fair manner (i.e., by assuming that published experimental
and theoretical results are correct if we cannot indicate any
specific mistake invalidating them), we arrive at the
conclusion that some of the concepts of statistical physics
commonly used for the theoretical description of
the interaction of fluctuating
fields with matter need to be reconsidered.

The structure of this paper is as follows. In Sec.~2 we briefly
discuss the Nernst theorem in the Lifshitz theory.
Sec.~3 is devoted to the same subject in application to the
proposed modifications of the Lifshitz theory. In Sec.~4 we consider
what the experiments say and if
they are reliable. Sec.~5 considers what
statistical physics says. Sec.~6 contains our conclusions.

\section{The Lifshitz theory and the Nernst heat theorem}

The Lifshitz theory provides an expression for the free energy
${\mathcal F}(a,T)$ of the fluctuating electromagnetic field
interacting with two thick uncharged plates (semispaces)
separated by a gap of width $a$ per unit area of plates.
It is supposed that this system is in thermal equilibrium at
temperature $T$. Material of the plates is described by the
dielectric permittivity $\varepsilon(\omega)$ depending only
on the frequency. Under these conditions ${\mathcal F}(a,T)$
is expressed in terms of the Fresnel reflection coefficients
$r_{\rm TM,TE}(i\xi_l,k_{\bot})$ for the transverse magnetic
(TM) and
 electric (TE) polarizations of the
electromagnetic field calculated at the imaginary Matsubara
frequencies $\xi_l=2\pi k_BTl/\hbar$, where $k_B$ is the
Boltzmann constant, $l=0,\,1,\,2,\,\ldots$, and
$\mbox{\boldmath$k$}_{\bot}=(k_x,k_y)$ is the projection of
the wave vector on the plane of the plates.\cite{2}

For materials with no free charge carriers (insulators)
$\varepsilon(i\xi)$ can be represented in the oscillator form
\begin{equation}
\varepsilon(i\xi)=1+\sum_{j=1}^{K}
\frac{g_j}{\omega_j^2+\xi^2+\gamma_j\xi},
\label{eq1}
\end{equation}
\noindent
where $\omega_j\neq 0$ are the oscillator frequencies and
$\varepsilon_0\equiv\varepsilon(0)<\infty$.
Electrons in metals are usually described by the Drude or plasma
models
\begin{equation}
\varepsilon_D(i\xi)=1+
\frac{\omega_p^2}{\xi(\xi+\gamma)},\qquad
\varepsilon_p(i\xi)=1+
\frac{\omega_p^2}{\xi^2},
\label{eq2}
\end{equation}
\noindent
where $\omega_p$ is the plasma frequency, $\gamma$ is the
relaxation parameter.

It was suggested\cite{9,10} to use the Nernst heat theorem as a test of
applicability of different models of $\varepsilon$ in the
Lifshitz theory. The entropy of the system under consideration
(the two plates interacting with the fluctuating field) per
unit area of plates is finite and can be calculated as
\begin{equation}
S_{\rm syst}(a,T)=
-\frac{\partial{\mathcal F}(a,T)}{\partial T}-
\frac{\partial{\mathcal F}_n(T)}{\partial T}\equiv
S(a,T)+S_n(T).
\label{eq4}
\end{equation}
\noindent
Here, $S(a,T)$ is the separation-dependent part of the
entropy related to the interaction between the fluctuating field
and the plates, and ${\mathcal F}_n\>(S_n)$ are the parts of the
free energy (entropy) of the system which do not depend on $a$.
The quantities ${\mathcal F}_n\>(S_n)$ are related to the
noninteracting case (specifically, they contain the large free
energy and entropy of remote plates) and do not contribute
to the Casimir force.

There are different formulations of the third law of
thermodynamics (the Nernst heat theorem) in the literature
(some of them are discussed in Ref.~[30]).
Below throughout the text we use only the
standard formulation from textbooks which is the
following.\cite{30,31} When $T\to 0$, the entropy of an equilibrium system
[in our case $AS_{\rm syst}(a,T)$ where $A$ is the area of the
plates] goes to a finite limit $S_{\rm syst,0}$ which does not
depend on volume, pressure, density or other thermodynamic
parameters of the system. According to quantum statistical
physics, we get\cite{30,31}
\begin{equation}
S_{\rm syst,0}=k_B\ln\,W_0,
\label{eq5}
\end{equation}
\noindent
where $W_0$ is an integer number describing the degree of
degeneracy of the ground state of the
system. If the ground state is nondegenerate, $W_0=1$,
one has $S_{\rm syst,0}=0$. The latter, however, is not
necessary to satisfy the Nernst theorem
as formulated above. It is important only
that $S_{\rm syst,0}$ does not depend on the
continuous thermodynamic
parameters, specifically, on $a$. Keeping in mind that
$S(a,T)$ is the part of the entropy depending on $a$,
the necessary requirement for the satisfaction of the Nernst
theorem is that $S(a,T)\to 0$ when $T\to 0$. If $S(a,T)$
goes to some function of $a$, $f(a)$, when $T\to 0$, the
Nernst theorem is violated because $f(a)$ cannot be compensated
by the $a$-independent limit of the quantity $S_n(T)$.

When $\varepsilon_D$ of the Drude model (\ref{eq2}) is
substituted into the Lifshitz formula for metals with perfect
crystal lattices, we get\cite{10}
\begin{equation}
S(a,0)=S_D(a,0)=-\frac{k_B\zeta(3)}{16\pi a^2}\left[
1-4\frac{\delta_0}{a}+12\left(\frac{\delta_0}{a}\right)^2-
\cdots\right]<0,
\label{eq6}
\end{equation}
\noindent
where $\delta_0=c/\omega_p$ is the skin depth and $\zeta(z)$ is
the Riemann zeta function.
A metal with perfect crystal lattice
is a truely equilibrium system. Thus in this case we deal with
the violation of the Nernst heat theorem. It was argued in the
literature\cite{27} that with the decrease of $T$ the frequency
region of the anomalous skin effect, where local description by
means of $\varepsilon_D(\omega)$ is inapplicable, extends to low
frequencies. This objection, however, does not solve the problem.
First, for any low $T$ there exists some narrow region of small
frequencies $[0,\omega_0]$ where local desctiption by means of
$\varepsilon_D(\omega)$ is applicable. Then the result (\ref{eq6})
remains valid because it originates from the zero-frequency
term of the Lifshitz formula. Second, the Drude model at low
$T$ was used for the interpolation between the regions of the
normal skin effect and infrared optics in the classical
theories by Bloch, Gr\"{u}neisen and Debye.\cite{32}
Although such a model approach does not provide an exact
description of real metals due to the existence of the anomalous
skin effect, it seems strange that it leads to the violation
of the Nernst theorem when used in combination with the
Lifshitz formula. Note that for metals with impurities the
Lifshitz formula combined with the Drude model satisfies
the Nernst theorem.\cite{8,32a}
This is a step forward in the resolution of the problem but
does not solve it because the introduction of impurities
might result in a violation of the thermal equilibrium which
for sure takes place for perfect crystal lattices
(see a discussion\cite{32b}).
At the same
time the substitution of $\varepsilon_p$ of the plasma
model (\ref{eq2}) into the Lifshitz formula leads to
$S_p(a,0)=0$.
For insulators it was shown\cite{16} that $S(a,T)$ calculated
with the dielectric permittivity (\ref{eq1}) goes to zero
when $T$ vanishes. If, however, the dc conductivity
$\sigma_0(T)$ is taken into account,
\begin{equation}
\varepsilon_{\rm dc}(i\xi,T)=\varepsilon(i\xi)+4\pi
\frac{\sigma_0(T)}{\xi},
\label{eq7}
\end{equation}
\noindent
it results in the violation of the Nernst theorem
\begin{equation}
S(a,0)=S_{\rm dc}(a,0)=\frac{k_B}{16\pi a^2}\left[
\zeta(3)-\mbox{Li}_3(r_0^2)\right]>0.
\label{eq8}
\end{equation}
\noindent
Here, $\mbox{Li}_3(z)$ is the polylogarithm function and
$r_0=(\varepsilon_0-1)/(\varepsilon_0+1)$.

To avoid the violation of the Nernst theorem and contradictions
with the experimental data (see Sec.~4) in numerous
applications of the Lifshitz theory, the following
phenomenological prescription was proposed.\cite{33,34}
When applying the Lifshitz theory to metals, conduction
electrons should be described by the plasma model. In the
application of this theory to dielectrics, dc conductivity
should be omitted. Keeping in mind that all materials
can be divided into metals (whose conductivity is not equal
to zero at $T=0$) and dielectrics (whose conductivity
vanishes when $T\to 0$), this prescription can be considered
as universally applicable.
In some sense it is not new because metals were often described
in the literature by means of the plasma model\cite{6,35}
and the dc conductivity of dielectrics was almost always
omitted.\cite{14} It was generally believed, however, that
with account of relaxation properties of conduction electrons
(i.e., using the Drude model) and of the dc conductivity
of dielectrics slightly more exact results would be obtained.
The new fact recognized in the last few years is that
the inclusion of these features leads to
drastically different calculational results which are in
conflict with thermodynamics and contradict  the
experimental data. This fact invites reconsideration of the
Lifshitz theory and careful analysis of all assumptions
laid in its foundation.

\section{The Nernst heat theorem in the modifications of the
Lifshitz theory}

The most general modification\cite{19} leaves the formalism
of the Lifshitz theory unchanged but replaces the Fresnel
reflection coefficients, $r_{\rm TM,TE}(i\xi_l,k_{\bot})$
with the modified ones, $\tilde{r}_{\rm TM,TE}(i\xi_l,k_{\bot})$,
which take into account both the drift and diffusion currents by
means of the Boltzmann transport equation.
The modified reflection coefficients depend on a new parameter
$\kappa$ which has the physical meaning of an inverse screening
radius. It is equal to $\kappa_{\rm DH}$ or $\kappa_{\rm TF}$ for
Debye-H\"{u}ckel and Thomas-Fermi screening radia applicable for
the Maxwell-Boltzmann and Fermi-Dirac statistics, respectively.
For dielectrics $(\kappa=\kappa_{\rm DH})$ at $\xi=0$ the
coefficient $\tilde{r}_{\rm TM}(0,k_{\bot})$ was first obtained
in Ref.~\refcite{18}. The modified reflection coefficients
$\tilde{r}_{\rm TM,TE}$  were also phenomenologically
expressed\cite{20} in terms of $\mbox{\boldmath$k$}$-dependent
dielectric permittivities in the random phase approximation
(recall that in the presence of a gap between semispaces the
translation invariance in space is violated and the nonlocal
dielectric permittivity $\varepsilon_z$ depending on
$\mbox{\boldmath$k$}$ does not exist as a rigorous
mathematical concept\cite{36}).

For metals with perfect crystal lattice by using
$\kappa=\kappa_{\rm TF}$ it was shown\cite{22,23} that the
modified entropy $\tilde{S}(a,0)=S_D(a,0)<0$, as can be
seen from Eq.~(\ref{eq6}). Thus, in this case the modification
of the Lifshitz theory proposed\cite{18}\cdash\cite{20} suffers
from the same thermodynamic difficulty as the standard Drude
model.
For dielectric materials $(\kappa=\kappa_{DH})$ the situation
turned out to be more involved. Under the condition that the
density of charge carriers $n(T)\to 0$ more quickly than
$T^{1+\alpha}$ with $\alpha>0$ (this is the case for intrinsic
semiconductors) it was shown\cite{21,24} that the modified
entropy $\tilde{S}(a,0)=0$, i.e., the Nernst theorem is
satisfied. In the two Comments\cite{25,26} it was stressed,
however, that for dielecric materials not satisfying
this condition (for instance, for doped semiconductors with
$n<n_{\rm cr}$, semimetals of dielectric type and solids
with ionic conductivity) the modified Lifshitz theory
violates the Nernst heat theorem. In this case it holds
$\tilde{S}(a,0)=S_{dc}(a,0)>0$ where $S_{dc}$ is defined
in Eq.~(\ref{eq8}). For dielectric materials under
consideration $n(T)$ does not go to zero with vanishing $T$ and
conductivity vanishes with temperature due to the vanishing
mobility of charge carriers.

The result that the modifications of the Lifshitz theory are
in disagreement with thermodynamics was disputed in the
literature. Thus, it was claimed\cite{29} that the approach
of Ref.~\refcite{19} satisfies the Nernst theorem for all
dielectrics. However, in the respective proof it was assumed that
$n(T)\to 0$ when $T\to 0$. The above-mentioned dielectric
materials for which this is not the case were not discussed.
Reply\cite{27} claimed that the materials leading to
conflicts with thermodynamics in the modified Lifshitz
theory are amorphous glass-like bodies which are out of
equilibrium state and have a big entropy at $T=0$. The Nernst
theorem is not valid for such bodies. It is true that glass-like
bodies must not satisfy the Nernst theorem.
The arguments in the Reply\cite{27} are, however, somewhat
contradictory. The point is that we consider not the entropy
 $S_n(T)$ of the plate made of a glass-like material
 (SiO$_2$ for instance\cite{14}), but the entropy of the
 interaction with the fluctuating field $S(a,T)$ (see Sec.~2).
 If the fluctuating field is in equilibrium with the plate
 (as is assumed in Ref.~\refcite{14}), one can apply the Lifshitz
 theory. In this case, however, in accordance with the
 Nernst theorem, $S(a,T)$ must vanish when $T$ vanishes.
 In fact the input data for the Lifshitz formula are the
 values of $\varepsilon(i\xi)$ which are quite similar for
 the amorphous and polycryctal SiO$_2$. The Lifshitz formula
 is applicable when the fluctuating field is in equilibrium
 with the material of the plate. This formula is incapable
 of distinguishing between the cases when the plate material is
 in equilibrium or out of equilibrium. Reply\cite{27} does
 not also provide a response concerning the existence of
 crystallic materials (semimetals of the dielectric type, for
 instance) leading to the violation of the Nernst theorem
 in the proposed modifications of the Lifshitz theory.

Both Replies\cite{27,28} cast doubts on the fact
that there are dielectric
materials for which $n(T)$ does not go to zero when $T\to 0$
with a reference to the measurements\cite{37}
for SiO$_2$ performed
in the region from 433\,K to 473\,K. Such high-$T$ results
seem to be irrelevant to the problem under consideration.
Independent measurements of all three parameters, conductivity,
$n$ and mobility, demonstrate\cite{38} that ``mobility has the
dominant influence upon the conductivity-temperature
dependence.'' As was recently confirmed,\cite{39}
``On long time scales the `mobile' ion density must be the
total ion concentration. This `long run' may be years or
more, and ions trapped for so long are for all practical purposes
immobile. Nevertheless, unless there are infinite barriers in
the solid, which is unphysical, in the very long run all ions
are equivalent.'' Thus, for ionic conductors (like amorphous
SiO$_2$) $n$ does not vanish when $T\to 0$. The same
conclusion holds for compensated semiconductors of the dielectric
type. If the density of donor atoms $n_d$ is larger than the
density of acceptor atoms $n_a$, the density of charge
carriers at low $T$, $n_d-n_a$, remains constant.\cite{40}
One more example is provided by semimetals of the dielectric
type which are crystal materials with a regular structure.
For these materials the Fermi energy is at a band where
the density of states is not equal to zero. The number of
charge carriers near the Fermi surface is fixed and determined
by the structure of the crystal lattice. For both compensated
semiconductors and semimetals of dielectric type conductivity
vanishes due to vanishing mobility.\cite{41,42}
All the above testifies that the problem of thermodynamic
inconsistency of the proposed modifications of the Lifshitz
theory deserves serious attention.

\section{What experiments say and is it reliable}

It was widely discussed in the literature that the
measurement data of the
experiments with a micromechanical oscillator\cite{11,12}
exclude the use of the Drude model for the calculation of
the thermal Casimir force between metals but are
consistent with the use of the plasma model.
The experiments with an atomic force microscope\cite{13}
and Bose-Einstein condensate\cite{14} are inconsistent
with the inclusion of the dc conductivity of a dielectric
plate but consistent with the theory omitting this
conductivity. These results are related to the standard
Lifshitz theory. They are obtained at a 99.9\% and 95\%
confidence levels with respect to experiments of
Refs.~\refcite{12,13} and at a 70\% confidence level
for the experiment of Ref.~\refcite{14}.

Just after the modifications of the Lifshitz theory were
proposed, the obtained theoretical results (which are
almost coincident for all three variants of the
modified theory) were compared with the experimental
data. For metals, it was found\cite{22,23,26} that
the experimental data\cite{12} exclude the modified
Lifshitz theory at a 99.9\% confidence level.
For dielectrics, the data of the experiment\cite{13}
exclude the predictions of the modified theory at a 70\%
confidence level.\cite{21,24,25} It was found also that
the data of the experiment\cite{14} determined at a 70\%
confidence level are not precise enough and do not
permit to make a conclusive comparison with theory.
The point is that it is consistent with both the
standard Lifshitz theory with dc conductivity
excluded and with the modified Lifshitz theory.

Note that it was claimed\cite{20} that the experimental
data\cite{13} can hardly distinguish between the standard
Lifshitz theory with omitted dc conductivity of
dielectric Si and the modified theory. This claim is
based on a complete misunderstanding of statistical
procedures used for the comparison between experiment
and theory. Thus, in Fig.~1a of Ref.~\refcite{20}
the experimental data are shown with errors determined
at a 70\% confidence level, but the width of the
theoretical band related to the modified theory was
calculated at a 95\% confidence level (i.e.,
artificially widened in order to make theory
consistent with the data). Such a comparison is
evidently irregular. In the Erratum,\cite{20}
instead of plotting the theoretical band at a 70\%
confidence level, the experimental errors were increased
by calculating them at a 95\% confidence level.
This is, however, meaningless because the data\cite{13}
are not of sufficient precision for the conclusive
comparison with the modified Lifshitz theory at a 95\%
confidence level.\cite{21} If a comparison at the 70\%
confidence level would be made, the result\cite{21}
on the exclusion of the modified theory is reproduced.

Thus, the Drude model approach and the modified
Lifshitz theory are in disagreement with the experimental
data. The question arises what is the reliability of
these experiments. The experiments under consideration were
repeated several times with the same result and the
most conservative statistical procedures for the data
processing and error analysis have been used. It was
claimed, however, that there is an anomalous distance
dependence of the gradient of the electric force,
used for calibration of the Casimir setup, between an Au
plate and an Au spherical lens of 30\,mm radius.\cite{43}
The respective contact potential was found to be
separation-dependent. On this basis it was suggested to perform
a reanalysis of the previous experiments mentioned above.
These doubts cast on previous experiments with small spheres of
about $100\,\mu$m radia are not justified. The reason is that
the contact potential in the experiments\cite{11,12} was
measured to be constant over a wide range of separations and
the standard force-distance dependence for the electric force
was observed, as predicted by classical electrodynamics.
The possible reason for the anomalous dependence observed\cite{43}
is deviation of the mechanically polished and ground surface of
the centimeter-size radius from a perfect spherical shape.\cite{44}
An attempt to avoid this conclusion using the capacitance
measurements at large separations\cite{45} was shown to be
based on incorrect computations.\cite{46}
Because of this, continuing claims that important systematic
effects have not been properly taken care of in the
electrostatic calibrations in previous experiments, in our
opinion, are unfair and cannot be considered as a scientific
argument against these experiments.
This does not mean that there is no need to look for systematic
effects which might be present in previously performed
experiments. It would be desirable, however, that such kind
investigations were performed in the experimental configurations
maximally similar to the original ones and were not based on
far-reaching extrapolations.

\section{What statistical physics says}

Classical statistical physics permits one to calculate the free
energy for two remote plates consisting of mobile quantum
charges interacting with the quantized electromagnetic field.
In doing so, photons and charges are supposed to be in
thermal equilibrium at temperature $T$.
The obtained free energy\cite{17} is equal to the one
calculated by using the Lifshitz formula combined with the
Drude model (i.e., equal to one half of the result valid for
ideal metal plates).

Another consequence of statistical physics is the
Bohr-van Leeuwen theorem which states that in classical
systems at thermal equilibrium matter decouples from the
transverse electromagnetic field. Recently it was shown\cite{47}
that this theorem is satisfied if and only if at large
separations the reflection coefficient $r_{\rm TE}(0,k_{\bot})$
of nonmagnetic materials is equal to zero leading to the same
result for the free energy as the Lifshitz formula combined with the
Drude model. Thus, in the classical limit (at large separations)
the Drude model approach finds support from the source side of
statistical physics although it has difficulties with respect to
the Nernst theorem and disagrees with the experimental data at
short separations.

It this situation it is useful to reformulate the problem in
an equivalent way. It was shown\cite{48} that large negative
temperature correction arising in the Drude model approach at
short separations can be described as the contribution of
eddy currents. The absence of this contribution in the
measurement data was interpreted in a way that it was
somehow reduced.\cite{48} The mechanism of this reduction
remains, however, unclear. As a possible resolution of
the problem the standard Planck distribution was
modified\cite{49} by including a phenomenological
parameter $D$ taking into account the ``saturation effects''.
In this way an agreement between the Lifshitz theory combined
with the Drude model and experimental data\cite{12}\cdash\cite{14}
was achieved. However, the relative arbitrariness in the
value of $D$ remains a problem.

The roots of the controversial situation under consideration
might be connected with the use of some basic statements
of statistical physics outside of their application region.
It is common knowledge that when a physical system deviates from
the equilibrium state (for instance, when a semiconductor is
placed in an external field) the fluctuation-dissipation
theorem is violated. In this respect it is pertinent to recall
that both the Lifshitz theory combined with the Drude model
and its modifications\cite{18}\cdash\cite{20} include
transport phenomena in an external field and, thus, violate the
applicability condition of the fluctuation-dissipation theorem
on which they are based. The possibility of such violation
is explicitly admitted by the statement\cite{18} that ``It is not
clear if the fields with the very low frequencies... are
in thermal equilibrium with bodies. The problem is worth
experimental investigation.'' In our opinion
experiments\cite{12}\cdash\cite{14} have already solved this
problem in the most unambiguous manner.

\section{Conclusions}

{}From the foregoing we arrive to the following conclusions.
\begin{enumerate}
\item[1)]
For metals with perfect crystal lattices the Lifshitz theory
combined with the Drude model violates the Nernst theorem.
The Nernst theorem is satisfied when the relaxation is
nonzero at zero temperature, i.e. when impurities are
taken into account.
The Lifshitz theory including the dc conductivity
of dielectrics and modifications of this
theory violate the Nernst theorem for wide classes of
different materials.
\item[2)]
The experimental data of several experiments are inconsistent
with the Lifshitz theory combined with the Drude model or
including the dc conductivity and with
the modifications of this theory.
Keeping in mind that the Drude relation correctly describes
the response of a metal to real (external) electric field,
the reason of this inconsistency might be connected with
some fundamental differences between real and fluctuating
fields.
\item[3)]
Phenomenologically, contradictions of the Lifshitz theory with
both the Nernst theorem and the experimental data disappear
if the free charge carriers are described by means of the
plasma model in metals and are disregarded in dielectrics.
Similar to any phenomenological approach, this one is
useful as a practical matter but cannot be offered as an
alternative to a complete theoretical description which
remains unknown.
\item[4)]
In our opinion,
there are concepts of statistical physics
related to the theoretical description of the interaction
of classical and quantum fluctuating fields with matter
that might need a reconsideration.
Opinions on this subject vary and the consensus is not yet
achieved.
\end{enumerate}

\section*{Acknowledgments}

The authors are grateful to the
Deutsche Forschungsgemeinschaft
Grant No.~GE\,696/9--1
for partial financial support.



\begin{thebibliography}{99}
\bibitem{1}
H.~B.~G.~Casimir,
{\em Proc. K. Ned. Akad. Wet.}
{\bf 51} 793 (1948).
\bibitem{2}
M.~Bordag, G.\ L.\ Klimchitskaya, U.\ Mo\-hi\-deen and
V.\ M.\ Mostepanenko,
{\it Advances in the Casimir Effect}
(Oxford University Press, Oxford, 2009).
\bibitem{3}
 G.\ L.\ Klim\-chits\-kaya, U.\ Mo\-hi\-deen and
V.\ M.\ Mos\-te\-pa\-nen\-ko,
{\em Rev. Mod. Phys.} {\bf 81}, N4 (2009);
ArXiv:0902.4022.
\bibitem{4}
M.~Bostr\"{o}m and B.~E.~Sernelius,
{\em Phys. Rev. Lett.} {\bf 84}, 4757 (2000).
\bibitem{5}
M.~Bordag, B.~Geyer, G.\ L.\ Klimchitskaya and
V.\ M.\ Mostepanenko,
{\em Phys. Rev. Lett.} {\bf 85}, 503 (2000).
\bibitem{6}
C.~Genet, A.~Lambrecht and S.\ Reynaud,
{\em Phys. Rev. A} {\bf 62}, 012110 (2000).
\bibitem{7}
J.~S.~H{\o}ye, I.~Brevik,  J.~B.~Aarseth and K.\ A.\ Milton,
{\em Phys. Rev. E} {\bf 67}, 056116 (2003);
I.~Brevik,  J.~B.~Aarseth, J.~S.~H{\o}ye and K.~A.~Milton,
{\em Phys. Rev. E} {\bf 71}, 056101 (2005).
\bibitem{8}
J.~S.~H{\o}ye, I.~Brevik, S.~A.~Ellingsen and J.~B.~Aarseth,
{\em Phys. Rev. E} {\bf 75}, 051127 (2007).
\bibitem{9}
V.~B.~Bezerra, G.~L.~Klimchitskaya and V.~M.~Mostepanenko,
{\em Phys. Rev. A} {\bf 66}, 062112 (2002);
G.~L.~Klimchitskaya, U.~Mohideen and V.~M.~Mostepanenko,
{\em J. Phys. A: Math. Theor.} {\bf 40}, (F)339 (2007).
\bibitem{10}
V.~B.~Bezerra, G.~L.~Klimchitskaya, V.~M.~Mostepanenko
and C.~Romero,
{\em Phys. Rev. A} {\bf 69}, 022119 (2004).
\bibitem{11}
R.~S.~Decca,  E.~Fischbach, G.~L.~Klimchitskaya, D.~E.~Krause,
D.~L\'opez and V.~M.~Mostepanenko,
{\em Phys. Rev. D} {\bf 68}, 116003 (2003);
R.~S.~Decca,  D.~L\'opez, E.~Fischbach, G.~L.~Klimchitskaya,
 D.~E.~Krause and V.~M.~Mostepanenko,
 {\em Ann. Phys. (N.Y.)} {\bf 318}, 37 (2005).
\bibitem{12}
R.~S.~Decca,  D.~L\'opez, E.~Fischbach, G.~L.~Klimchitskaya,
 D.~E.~Krause and V.~M.~Mostepanenko,
{\em Phys. Rev. D} {\bf 75}, 077101 (2007);
{\em Eur. Phys. J. C} {\bf 51}, 963 (2007).
\bibitem{13}
F.~Chen,   G.~L.~Klimchitskaya,
V.\ M.\ Mos\-te\-pa\-nen\-ko and U.~Mo\-hi\-deen,
{\em Optics Express} {\bf 15}, 4823 (2007);
{\em Phys. Rev. B} {\bf 76}, 035338 (2007).
\bibitem{14}
J.~M.~Obrecht, R.~J.~Wild, M.~Antezza, L.~P.~Pitaevskii,
S.\ Stringari and E.~A.~Cornell,
{\em Phys. Rev. Lett.} {\bf 98}, 063201 (2007).
\bibitem{15}
G.~L.~Klimchitskaya and V.~M.~Mostepanenko,
{\em J. Phys. A: Math. Theor.} {\bf 41}, (F)312002 (2008).
\bibitem{16}
B.~Geyer, G.~L.~Klimchitskaya and V.~M.~Mostepanenko,
{\em Phys. Rev. D} {\bf 72}, 085009 (2005).
\bibitem{17}
P.~R.~Buenzli and Ph.~A.~Martin,
{\em Phys. Rev. E} {\bf 77}, 011114 (2008);
B.~Jancovici and L.~\v{S}amaj,
{\it Europhys. Lett.} {\bf 72}, 35 (2005).
\bibitem{18}
L.~P.~Pitaevskii,
{\it Phys. Rev. Lett.} {\bf 101}, 163202 (2008).
\bibitem{19}
D.~A.~R.~Dalvit and S.~K.~Lamoreaux,
{\it Phys. Rev. Lett.} {\bf 101}, 163203 (2008).
\bibitem{20}
V.~B.~Svetovoy,
{\it Phys. Rev. Lett.} {\bf 101}, 163603 (2008);
{\bf 102}, (E)219903 (2009).
\bibitem{21}
G.~L.~Klimchitskaya, U.~Mohideen and V.~M.~Mostepanenko,
{\em J. Phys. A: Math. Theor.} {\bf 41}, (F)432001 (2008).
\bibitem{22}
V.\ M.\ Mostepanenko, R.~S.~Decca, E.\ Fischbach, B.\ Geyer,
G.\ L.\ Klimchitskaya, D.\ E.\ Krause, D.\ L\'opez
and U.\ Mohideen,
{\it Int. J. Mod. Phys. A} {\bf 24}, 1721 (2009).
\bibitem{23}
V.\ M.\ Mostepanenko,
{\em J. Phys.: Conf. Series} {\bf 161}, 012003 (2009).
\bibitem{24}
G.~L.~Klimchitskaya,
{\em J. Phys.: Conf. Series} {\bf 161}, 012002 (2009).
\bibitem{25}
B.~Geyer, G.~L.~Klimchitskaya, U.~Mohideen and
V.~M.~Mostepanenko,
{\em Phys. Rev. Lett.} {\bf 102}, 189301 (2009).
\bibitem{26}
R.~S.~Decca,  E.~Fischbach, B.~Geyer, G.~L.~Klimchitskaya,
D.~E.~Krause, D.~L\'opez, U.~Mohideen and V.~M.~Mostepanenko,
{\em Phys. Rev. Lett.} {\bf 102}, 189303 (2009).
\bibitem{27}
L.~P.~Pitaevskii,
{\em Phys. Rev. Lett.} {\bf 102}, 189302 (2009).
\bibitem{28}
D.~A.~R.~Dalvit and S.~K.~Lamoreaux,
{\em Phys. Rev. Lett.} {\bf 102}, 189304 (2009).
\bibitem{29}
D.~A.~R.~Dalvit and S.~K.~Lamoreaux,
{\em J. Phys.: Conf. Series} {\bf 161}, 012009 (2009).
\bibitem{29a}
M.~Aizenman and E.~H.~Lieb,
{\em J. Stat. Phys.} {\bf 24}, 279 (1981).
\bibitem{30}
D.~Kondepugi and I.~Prigogine,
{\it Modern Thermodynamics}
(John Wiley \& Sons, New York, 1998).
\bibitem{31}
Yu.~B.~Rumer and M.~S.~Ryvkin,
{\it Thermodynamics,
Statistical Physics and Kinetics}
(Mir, Moscow, 1980).
\bibitem{32}
N.~W.~Ashcroft and N.~D.~Mermin,
{\em Solid State Physics}
(Saunders Col\-le\-ge, Philadelphia, 1976).
\bibitem{32a}
I.~Brevik,  S.~A.~Ellingsen, J.~S.~H{\o}ye and K.~A.~Milton,
{\em J. Phys. A: Math. Theor.} {\bf 41}, 164017 (2008).
\bibitem{32b}
G.~L.~Klimchitskaya and V.~M.~Mostepanenko,
{\em Phys. Rev. E} {\bf 77}, 023101 (2008);
J.~S.~H{\o}ye, I.~Brevik, S.~A.~Ellingsen and J.~B.~Aarseth,
{\em Phys. Rev. E} {\bf 77}, 023102 (2008).
\bibitem{33}
V.~M.~Mostepanenko and B.~Geyer,
{\it J. Phys. A: Math. Theor.} {\bf 41}, 164014 (2008).
\bibitem{34}
G.~L.~Klimchitskaya and B.~Geyer,
{\em J. Phys. A: Math. Theor.} {\bf 41}, 164032 (2008).
\bibitem{35}
J.~Schwinger, L.~L.~DeRaad and K.~A.~Milton,
{\em Ann. Phys. (N.Y.)} {\bf 115}, 1 (1978).
\bibitem{36}
G.~L.~Klimchitskaya and V.~M.~Mostepanenko,
{\it Phys. Rev. B} {\bf 75}, 036101 (2007);
B.\ E.\ Sernelius,
{\it Phys. Rev. B} {\bf 75}, 036102 (2007).
\bibitem{37}
M.~Tomazawa and D.-W.~Shin,
{\em J. Non-Cryst. Solids} {\bf 241}, 140 (1998).
\bibitem{38}
H.~J.~Sch\"{u}tt and E.~Gerdes,
{\em J. Non-Cryst. Solids} {\bf 144}, 14 (1992).
\bibitem{39}
J.~C.~Dure, P.~Maass, B.~Roling and D.~L.~Sidebottom,
{\em Rep. Progr. Phys.} {\bf 72}, 046501 (2009).
\bibitem{40}
K.~Seeger,
{\it Semiconductor Physics}
(Springer, New York, 1973).
\bibitem{41}
N.~F.~Mott,
{\em Phil. Mag.} {\bf 19}, 835 (1969).
\bibitem{42}
N.~F.~Mott,
{\it Metal-Insulator Transitions}, 2nd edition (Taylor and Francis,
London, 1990).
\bibitem{43}
W.~J.~Kim, M.~Brown-Hayes, D.~A.~R.~Dalvit, J.~H.~Brownell
and R.~Onofrio,
{\em Phys. Rev. A} {\bf 78}, (R)020101 (2008).
\bibitem{44}
R.~S.~Decca,  E.~Fischbach,  G.~L.~Klimchitskaya,
D.~E.~Krause, D.~L\'opez, U.~Mohideen and V.~M.~Mostepanenko,
{\em Phys. Rev. A} {\bf 79}, 026101 (2009).
\bibitem{45}
W.~J.~Kim, M.~Brown-Hayes, D.~A.~R.~Dalvit, J.~H.~Brownell
and R.~Onofrio,
{\em Phys. Rev. A} {\bf 79}, 026102 (2009).
\bibitem{46}
R.~S.~Decca,  E.~Fischbach, B.~Geyer, G.~L.~Klimchitskaya,
D.~E.~Krause, D.~L\'opez, U.~Mohideen and V.~M.~Mostepanenko,
ArXiv:0904.4720.
\bibitem{47}
G.~Bimonte,
{\em Phys. Rev. A} {\bf 79}, 042107 (2009).
\bibitem{48}
F.~Intravaia and C.~Henkel,
{\it Phys. Rev. Lett.} {\bf 103}, 130405 (2009).
\bibitem{49}
B.~E.~Sernelius,
{\em Europhys. Lett.} {\bf 87}, 14004 (2009);
{\em Phys. Rev. A} {\bf 80}, 043828 (2009).

\end{thebibliography}
\end{document}